\documentclass[aps,showpacs,amsmath,nofootinbib,amssymb,pra,superscriptaddress,preprint,floatfix]{revtex4-1}
\usepackage{graphicx}
\usepackage{amsmath}
\usepackage{hyperref}
\usepackage{dcolumn}
\usepackage{bm}
\usepackage{color}

\begin{document}

\title{Dynamics of Local Symmetry Correlators for Interacting Many-Particle Systems}

\author{P.~Schmelcher}
\email[]{pschmelc@physnet.uni-hamburg.de}
\affiliation{Zentrum f\"ur Optische Quantentechnologien, Universit\"{a}t Hamburg, Luruper Chaussee 149, 22761 Hamburg, Germany}
\affiliation{Hamburg Centre for Ultrafast Imaging, Universit\"{a}t Hamburg,
Luruper Chaussee 149, 22761 Hamburg, Germany}

\author{S.~Kr\"onke}
\affiliation{Zentrum f\"ur Optische Quantentechnologien, Universit\"{a}t Hamburg, Luruper Chaussee 149, 22761 Hamburg, Germany}
\affiliation{Hamburg Centre for Ultrafast Imaging, Universit\"{a}t Hamburg,
Luruper Chaussee 149, 22761 Hamburg, Germany}

\author{F.~K.~Diakonos}
\affiliation{Department of Physics, University of Athens, GR-15771 Athens, Greece}

\date{\today}

\begin{abstract}
Recently (PRL {\bf{113}}, 050403 (2014)) the concept of local symmetries in one-dimensional stationary wave propagation has been shown to lead to
a class of invariant two-point currents that allow to generalize the parity and Bloch theorem. In the present work we establish
the theoretical framework of local symmetries for higher-dimensional interacting many-body systems. Based on the 
Bogoliubov-Born-Green-Kirkwood-Yvon (BBGKY) hierarchy we derive the equations of motion of local symmetry correlators 
which are off-diagonal elements of the reduced one-body density matrix at symmetry related positions.
The natural orbital representation yields equations of motion for the convex sum of the local symmetry correlators of the natural orbitals
as well as for the local symmetry correlators of the individual orbitals themselves. An alternative integral representation with a unique interpretation
is provided. We discuss special cases, such as the bosonic and fermionic mean field theory and show in particular that the
invariant two-point currents are recovered in the case of the non-interacting one-dimensional stationary wave propagation.
Finally we derive the equations of motion for anomalous local symmetry correlators which indicate the breaking of a global
into a local symmetry in the stationary non-interacting case.
\end{abstract}


\maketitle

\section{Introduction}

Symmetries represent a cornerstone of modern quantum physics including many different fields, such as atomic, molecular 
or condensed matter physics \cite{Zare,Hammermesh,Harris,Dresselhaus}. As a consequence the corresponding mathematical
framework is very well developed 
and provides many powerful tools in order to obtain statements on the properties of specific systems even without
solving the underlying equations of motion (EOM). Examples are the discrete and continuous groups constituted by inversion, reflection, 
translation or rotation transforms which leave the underlying Hamiltonian invariant. Theorems deduced from the existence of these global symmetries,
such as the parity or Bloch theorem, are a key ingredient for all further analysis. The presence of symmetries has immediate consequences for
many properties of quantum systems, such as the existence of conserved quantum numbers or resulting selection rules for e.g. electromagnetic transitions
\cite{Hammermesh,Harris,Dresselhaus}.

The situation becomes much less transparent and accessible if a system does not exhibit a global symmetry which holds in
complete space but possesses different symmetries in different domains of space. This occurs typically for materials of a higher
degree of complexity ranging from self-assembled configurations of atoms on surfaces \cite{Rahsepar} to frustrated crystallization following a
liquid-glass transition \cite{Shintani}. Local order and local symmetries are ubiquitous also in larger molecules or
even disordered matter \cite{Wochner2009}. One way to detect local order on the atomic scale uses X-ray cross correlation analysis
\cite{Wochner2009,Altarelli,Wochner2011,Lehmkuehler}. In view of the existing powerful methodology to tackle global symmetries
the question arises whether the presence of local symmetries also allows for the development of a predictive formalism and corresponding
mathematical framework. This is a question of major importance not only for principal and fundamental reasons but in particular since 
materials with local symmetries provide a bridge between ideal crystals and disordered matter and it is to be expected that they
will lead to a rich and unique phenomenology. 

\begin{figure}[t!]
\centering
\includegraphics[width=.65\columnwidth]{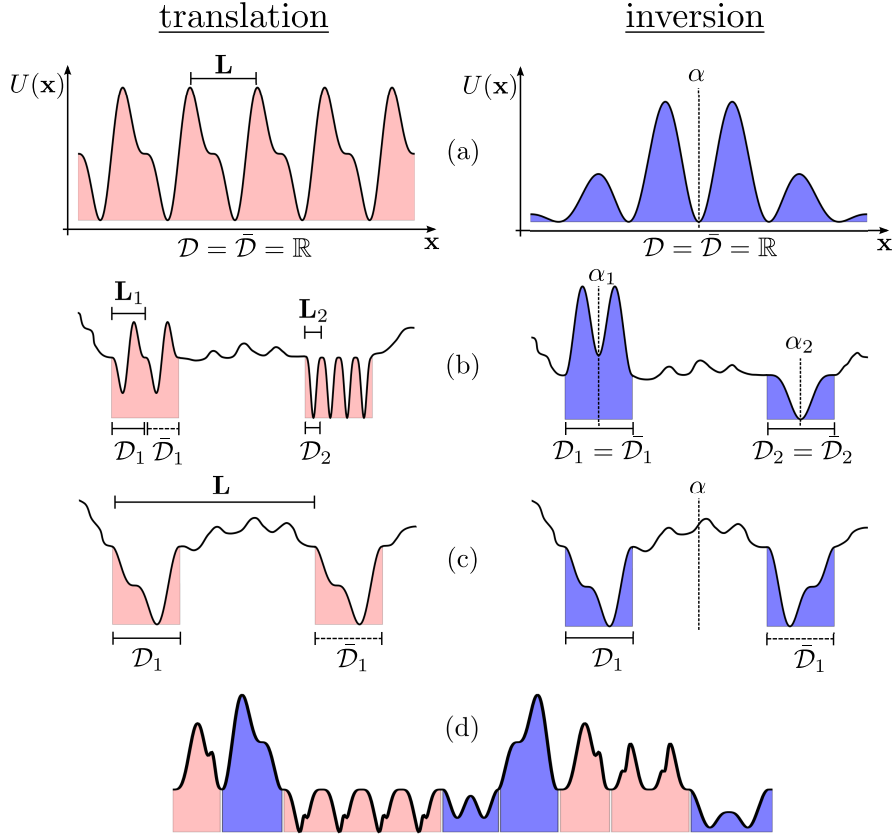}
\caption{
(Color online)
Illustration of different types of global and local symmetries, distinguishing between 
translation by $L$ or inversion through $\alpha$. Each symmetry maps a domain $\mathcal{D}$ to $\bar{\mathcal{D}}$: 
(a) global symmetry, (b) nongapped local symmetry, (c) gapped local symmetry, (d) complete local symmetry.
Note that $\mathcal{D}$ is mapped onto itself for a nongapped inversion symmetry.}
\label{fig1}
\end{figure}

Recently several steps in the direction of a theory of local symmetries have been performed indeed 
\cite{Kalozoumis2013a,Kalozoumis2014a,Kalozoumis2014b,Kalozoumis2013b,Kalozoumis2015a,Morfonios2014,
Kalozoumis2015b,Zambetakis2016,Kalozoumis2016a,Wulf2016}. The focus was hereby on the
case of non-interacting stationary one-dimensional single particle or wave mechanical (acoustics, optics) problems
which are described by the corresponding Schr\"odinger or Helmholtz
equation $\Psi''(x)+S(x)\Psi(x)=0$ where the prime denotes differentiation with respect to the spatial variable $x$.
Here $S(x)=\kappa^2(x)$ is a real function generated by an effective wave vector $\kappa(x)$,
which describes the inhomogeneity of the medium where the wave propagates.
For a matter wave $\Psi(x)$, $S(x)=\frac{2 m}{\hbar^2}(E-U(x))$ is the scaled kinetic energy of a particle with mass $m$ 
and energy $E$ which moves in a potential $U(x)$.
The complexity arises due to the presence of the potential term, which can exhibit a plethora of different local symmetries. Here, each local 
symmetry is associated with a local coordinate transform $F:\mathcal{D}\rightarrow\bar{\mathcal{D}}$, $x\mapsto 
F(x)$, which leaves the potential invariant: $U(F(x))=U(x)$. In what follows, we will focus on $F(x)=\sigma x+L$, i.e.\ 
on local discrete inversion (parity) symmetries ($\sigma=-1$) and local translation symmetries ($\sigma=1$). An illustration of such a 
composite potential landscape is provided in Figure \ref{fig1}. As it was recently found \cite{Kalozoumis2013b,Kalozoumis2014a} local symmetries
lead to the existence of invariant two-point correlators (ITPC) which possess the dimensionality of currents and are constant in the
corresponding domains of local symmetry. Two types of these ITPC were explored \cite{Kalozoumis2014a} and read as follows

\begin{equation}
Q =  \frac{1}{2i} \left[ \sigma {\Psi}(x) {\Psi}'(y) - {\Psi}(y) {\Psi}'(x) \right] = const. \hspace*{0.8cm} \forall x \in \mathcal{D}
\label{Q1}
\end{equation}
\begin{equation}
\widetilde{Q} = \frac{1}{2i} \left[ \sigma {\Psi}^*(x) {\Psi}'(y) - {\Psi}(y) {\Psi}'^*(x) \right] = const. \hspace*{0.8cm} \forall x \in 
\mathcal{D}
\label{Q2}
\end{equation}

\noindent
where $y=F(x)$ lies in the co-domain $\mathcal{\bar{D}} = F(\mathcal{D})$ of the local symmetry (see figure \ref{fig1}).
$\Psi$ is the wave function i.e. a solution of the stationary Schr\"odinger equation and the star denotes complex
conjugation. The above quantities are defined for each domain $\mathcal{D}$ of local symmetry separately and when considering complete real
space and a complete local symmetry (see Figure 1(d)) they are piecewise constant.
Looking back at the special case of a global symmetry which is respected also by the corresponding boundary conditions,
one can show \cite{Kalozoumis2014a} that $Q$ vanishes and $\frac{\widetilde{Q}}{J}$ represents the eigenvalue (parity or Bloch phase)
under the symmetry operation applied to the wave field, where $J$ is the well-known globally conserved probability current
\begin{equation}
J = \frac{1}{2i}\left[\Psi^*(x) \Psi'(x) - \Psi(x) \Psi'^*(x)\right]
\label{J}
\end{equation}

\noindent
Therefore, finite values of $Q$ indicate the symmetry breaking if we convert a global to a local symmetry, 
valid only in a limited spatial domain. This leads \cite{Kalozoumis2014a} to the 
generalization of the parity and Bloch theorem in the presence of local symmetries

\begin{equation}
\Psi(y)= \Psi(F(x))=\frac{1}{J} \left[ \widetilde{Q} \Psi(x) - Q \Psi^*(x) \right] \hspace*{0.8cm} \forall x \in \mathcal{D}
\label{genpbt}
\end{equation}

\noindent
The above relation (\ref{genpbt}) provides the mapping of the wave field between a spatial point $x$ and its symmetry transform $y$
which is solely determined by the ITPC $Q,\widetilde{Q}$. We remark that the above considerations can be extended beyond the
linear transforms of inversion and translation symmetry \cite{Zambetakis2016}.

The ITPC are the key quantities characterizing local discrete symmetries in one spatial dimension as has been shown in several works
\cite{Kalozoumis2014b,Kalozoumis2013b,Kalozoumis2015a,Morfonios2014,Kalozoumis2015b,Zambetakis2016,Kalozoumis2016a,Wulf2016}.
Applications to photonic multilayers have demonstrated that perfectly transmitting resonances can be completely classified according to
sum rules imposed on the ITPC \cite{Kalozoumis2013b}. Even more, a construction principle has been derived which allows for a systematic design of
locally symmetric multilayered structures that exhibit perfectly transmitting resonances at desired energies \cite{Kalozoumis2013b}.
Very recently a local basis approach has been developed \cite{Zambetakis2016} which provides a classification of the solutions
of the wave equation in terms of ITPC and consequently a computational strategy for the determination of the wave field in complex
locally symmetric potential landscapes (see Figure \ref{fig1}). Further very recent extensions of the theory of ITPC include 
time-dependent periodically driven \cite{Wulf2016} and parity-time symmetric setups with gain and loss \cite{Kalozoumis2014b,Kalozoumis2016a}
where e.g. the ITPC provide an order parameter for the phase diagram of the scattering states \cite{Kalozoumis2014b}. 
The first experimental detection of the ITPC has been achieved in acoustic waveguides \cite{Kalozoumis2015b} where both the phase and magnitude of the
pressure field yield a complete reconstruction of the complex ITPC. Here the theory has been extended to include losses and
the experiment has also verified the occurence of perfectly transmitting resonances \cite{Kalozoumis2015b}.

While the above-mentioned theoretical framework and applications focus on one-dimensional stationary and non-interacting wave mechanical systems
as ideally realized in acoustics and optics we develop here the theoretical framework for local symmetries (local inversion and local translation)
in interacting many-body systems including higher dimensions and dynamics.
The key idea is to define the (canonical) local symmetry 
correlators of an interacting many-body system as coherences of the reduced 
density matrices (RDM) between local symmetry related points in space. Starting 
from the well-known Bogoliubov-Born-Green-Kirkwood-Yvon (BBGKY) hierarchy of 
equations of motion for the RDM \cite{Born,Bogoliubov,Kirkwood,Yvon,Bonitz,Akbari} and focusing on the dynamics of the 
one-body RDM local symmetry correlators, we arrive at a continuity-like 
equation for the latter, involving the divergence of non-local two-point 
current densities and a local source term stemming from the interaction with 
the surrounding particles. These equations become more transparent if one 
employs the diagonal natural orbital representation of the one-body RDM such 
that the local symmetry correlators and their corresponding non-local
current densities are represented as convex sums over local symmetry correlators 
of the individual natural orbitals and their respective non-local current 
densities. Corresponding, continuity-like equations of motion for the 
local symmetry correlators of the individual natural orbitals are derived. 
Moreover, an integral representation for the equations of motion for the 
one-body local symmetry correlators is presented, which allows for a novel 
interpretation of the interplay between non-local currents and 
local symmetry correlators. Furthermore, we discuss the impact of the 
range of the interaction potential on the local source term of the derived 
continuity-like equations of motion.
This provides us with the fundamental framework of local symmetries of interacting many-body systems be it
fermions or bosons. Several special cases are subsequently discussed and it is, in particular, shown that the previously
developed non-interacting theory in one spatial dimension is indeed contained in the current general approach.
Besides the above we also establish the EOM for the anomalous symmetry correlators specializing to the
ITPC $Q$ for the non-interacting case.

\noindent
This work is structured as follows. Section II contains the theoretical framework. It introduces our 
Hamiltonian of the interacting particle system and some important properties of the resulting density matrix.
The BBGKY hierarchy, being the starting-point of our theoretical investigation, is subsequently discussed and
we review the EOM for the natural orbitals and their populations. In section III the
EOM for the symmetry correlators are derived and discussed in particular in the presence of local symmetries.
Different representations are shown and special cases, such as the non-interacting situation as well as
mean field theories of bosons and fermions are addressed. The EOM for the anomalous symmetry correlator 
reducing to the known symmetry breaking two-point current in the stationary one-dimensional and non-interacting case are
established as well, and the uniqueness of this construction is critically examined. Finally, in section IV we conclude and provide some outlook.

\section{Theoretical Framework}

As described above significant progress has been achieved in terms of both the formal description of local symmetries
and their impact and applications. Still, the focus so far is on a one-dimensional stationary non-interacting theory, which nicely applies
to acoustics and optics and non-interacting i.e. single particle quantum mechanics. Therefore the obvious question arises
whether the theoretical concept of local symmetries can be extended to the time-dependent situation, to higher dimensions
and interacting many-body systems.
According to observation, local symmetries and local order do in many cases occur (in nature) when many interacting degrees
of freedom cooperate and a high degree of complexity or nonequilibrium is met. In view of this the pressing question arises 
what an adequate level of description for the local symmetries of interacting systems would be. Trying to straightforwardly
generalize the above-mentioned approach of the non-interacting case but now starting with the $N$-particle Schr\"odinger equation
including external (locally symmetric) potentials and two-body interactions and following analogous steps in the derivation leads to a conflicting
situation. While in the non-interacting case a particle feels only the external local potential $U(x)$ and it therefore definitely belongs
to a certain domain of a given local symmetry, the case of interacting particles leads to the situation that two
interacting particles are in general located in two different local symmetry domains. As a consequence, the reduction and derivation
of expressions similar to eq.(\ref{Q1},\ref{Q2},\ref{genpbt}) are impossible. Even if one assumes sufficiently short-ranged (or contact)
interactions, such that ideally two-interacting particles always belong to the same domain of local symmetry, the resulting 
condition represents a $d \cdot N$-dimensional divergence of a $N$-dimensional generalization of the ITPC in eqs.(\ref{Q1},\ref{Q2})
in $d$ spatial dimensions. This condition cannot be easily used to advance, and, in particular, it cannot be straightforwardly integrated as it is
the case for the non-interacting one-dimensional theory.

In view of the above an approach based on integrated effective degrees of freedom is desirable. Such an approach is provided by
the corresponding density matrices and their EOM, the BBGKY hierarchy \cite{Born,Bogoliubov,Kirkwood,Yvon,Bonitz,Akbari}.
In the following we will provide
our setup and Hamiltonian, as well as important properties of the RDM and their equations motion. This will be the basis
for our derivation of the EOM of the symmetry correlators in the presence of local symmetries.

\subsection{Hamiltonian and Reduced Density Matrices for Interacting Particle Systems}

Our starting-point is a system of $N$ identical particles governed by the Hamiltonian

\begin{equation}
{\cal{H}} = \sum_{i=1}^{N} \left(-\frac{1}{2} \Delta_i + U(\bf{x}_i) \right) + \frac{1}{2} \sum_{i \neq j} V({\bf{x}}_i,{\bf{x}}_j)
\label{Hamiltonian}
\end{equation}

\noindent
where we have assumed atomic units (a.u.) and ${\bf{x}}$ are space-spin coordinates. $U({\bf{x}})$ is the single particle potential
which we will, later on, assume to possess local symmetries. $V({\bf{x}}_i,{\bf{x}}_j)$ represents the interaction between the
$i$-th and $j$-th particle. Solutions to the time-dependent Schr\"odinger equation $i \partial_t \Psi = {\cal{H}} \Psi$ are then
given by the time-dependent wave packets $\Psi({\bf{x}}_1,...,{\bf{x}}_N,t)$ whose $n-$th order ($n-$body) RDM are given by

\begin{equation}
\rho^{(n)} (X_n;X_n^{\prime};t)= \int \Psi^{*} (X_n^{\prime},X_n^c,t) \Psi (X_n,X_n^c,t) dX_n^c
\end{equation}

\noindent
with the compact notation $X_n=({\bf{x}}_1,...,{\bf{x}}_n)$ and $X_n^c=({\bf{x}}_{n+1},...,{\bf{x}}_N)$ where $c$ indicates
the complement w.r.t. the complete vector $X_N$. Here $dX_n^c = d{\bf{x}}_{n+1} ... d{\bf{x}}_N$ and the integration
w.r.t. ${\bf{x}}$ includes a spatial integration and a summation over the spin degrees of freedom. Let us mention some
relevant properties of the RDM. Of course, RDM inherit the bosonic or fermionic exchange symmetry 
within each of their two sets of coordinates from the underlying complete wave functions.
They obey a partial trace relation meaning that lower order RDM can be directly
calculated from a given order RDM by setting equal and integrating out the corresponding coordinates. RDM are hermitian and
possess (in our probabilistic state normalization) trace one. RDM are positive semidefinite yielding that all of their eigenvalues 
are equal to or greater than zero. The corresponding eigenvalue relation reads

\begin{equation}
\int \rho^{(n)}(X_n;X_n^{\prime};t) \phi_i^{(n)}(X_n^{\prime},t) dX_n^{\prime} = \frac{(N-n)!}{N!} \lambda_i^{(n)}(t) \phi_i^{(n)}(X_n,t)
\label{eigenvalueeq}
\end{equation}

\noindent
The eigenvectors $\phi_i^{(1)}$ and eigenvalues $\lambda_i^{(1)}$ of $\rho^{(1)}$ are refered to as the natural orbitals and their
occupation numbers or natural populations.
For bosonic systems, we have $0 \le \lambda_i^{(1)} \le  N$ whereas for fermions $0 \le \lambda_i^{(1)} \le 1$.
Specifically the diagonal representation of $\rho^{(1)}$ reads

\begin{equation}
\rho^{(1)} ({\bf{x}}_1;{\bf{x}}^{\prime}_1;t) = \frac{1}{N} \sum_{i} \lambda_i(t) \phi_i({\bf{x}}_1,t) \phi_i^{*}({\bf{x}}^{\prime}_1,t)
\label{rho1diag}
\end{equation}

\noindent
where, for simplicity, we have omitted the upper index $(1)$ indicating the order for the natural orbital $\phi_i$ and their
occupation numbers $\lambda_i$. 



\subsection{The BBGKY Hierarchy and the Equations of Motion for Natural Orbitals} \label{BBGKYsec}

Since the Schr\"odinger equation possesses a unique solution for a given initial wave function $\Psi(X_N;t=0)$ this holds also
for the EOM of the RDM. Working with the RDM  however has the advantage that typical observables are of one- or two-body character and
therefore the one-body or two-body RDM are sufficient to determine their expectation values. In essence, employing the $n$-th order RDM reduces
the number of relevant degrees of freedom enormously if $n<<N$ which is typically the case for many-body systems of identical 
particles. Since the procedure of deriving the BBGKY hierarchy will be applied also later in order to obtain the equation of motions
for the symmetry correlators which represent the analogue of the 
ITPC for interacting many-particle system (see eq.(\ref{Q1},\ref{Q2})), let us briefly summarize
the main steps. The BBGKY hierarchy is derived by taking the time derivative of the RDM $\rho^{(n)}$ of arbitrary degree $n$, employing the
underlying time-dependent Schr\"odinger equation, and dividing the total Hamiltonian ${\cal{H}}_{1....N}$ into
${\cal{H}}_{1....n}$ and ${\cal{H}}_{n+1,...,N}$ as well as coupling terms which emerge from the two-body interaction.
Here ${1....n}$ indicate the degrees of freedom of $\rho^{(n)}$ whereas ${n+1,...,N}$ are the ones traced out to obtain
$\rho^{(n)}$. Employing hermiticity of the Hamiltonian ${\cal{H}}_{n+1,...,N}$ and particle exchange symmetry of
either purely bosonic or fermionic wave functions one arrives at the hierarchy of EOM

\begin{eqnarray}
i \partial_t \rho^{(n)} (X_n;X_n^{\prime};t) =
\left( {\cal{H}}_{1...n} - {\cal{H}}_{1^{\prime}...n^{\prime}} \right) \rho^{(n)} (X_n;X_n^{\prime};t) \\ \nonumber
+ (N-n) \sum_{i=1}^{n} \int \left( V({\bf{x}}_i , {\bf{x}}_{n+1}) - V({\bf{x}}_i^{\prime} , {\bf{x}}_{n+1})  \right)
\rho^{(n+1)} (X_n,{\bf{x}}_{n+1};X_n^{\prime},{\bf{x}}_{n+1};t) d{\bf{x}}_{n+1}
\label{BBGKY}
\end{eqnarray}

\noindent
The so-called collision integral on the r.h.s. of this EOM stems from the two-body interaction coupling the two
above-mentioned subspaces. Obviously this is a system of $N$ coupled equations and constitutes a hierarchy in the
sense that every equation for $\rho^{(n)}$ couples via the integral expression to the next order $\rho^{(n+1)}$.
Solving this hierarchy, in particular while keeping the advantage $n<<N$ has to introduce a cutoff at comparatively low order.
At the same time a functional approximation to $\rho^{(n+1)}$ typically involving the lower order RDM has to be employed.
For the reconstruction of $\rho^{(n+1)}$ as a functional of lower order RDM
as well as for relevant properties of the truncated hierarchy of equations, such as particle number and energy
conservation as well as compatibility, we refer the reader to \cite{Bonitz,Akbari} and references therein. 
These features are not in the focus of the present work which aims at establishing the fundamental formalism
in the presence of local symmetries. The resulting framework can be employed both in the sense of a 
propagation method (if an appropriate closure approximation for the hierarchy is
utilized) and for obtaining analytical insights into the interplay between 
local symmetries and interactions.

It is instructive to inspect the lowest order equation separately which takes on the following appearance

\begin{align}
i \partial_t \rho^{(1)} ({\bf{x}}_1;{\bf{x}}_1^{\prime};t) =
\left( {\cal{H}}_{1} - {\cal{H}}_{1^{\prime}} \right) \rho^{(1)} ({\bf{x}}_1;{\bf{x}}_1^{\prime};t) \nonumber \\
+ (N-1) \int \left( V({\bf{x}}_1 , {\bf{x}}_{2}) - V({\bf{x}}_1^{\prime} , {\bf{x}}_{2})  \right)
\rho^{(2)} ({\bf{x}}_1,{\bf{x}}_{2};{\bf{x}}_1^{\prime},{\bf{x}}_{2};t) d{\bf{x}}_{2} 
\label{rho1eom}
\end{align}

\noindent
where ${\cal{H}}_1$ is the single particle part of the total Hamiltonian.
Focusing on ${\bf{x}}_1 = {\bf{x}}_1^{\prime}$ one can derive the standard continuity equation
for the particle density $n({\bf{x}},t) = \rho^{(1)} ({\bf{x}};{\bf{x}};t)$ which reflects particle number conservation

\begin{equation}
- \nabla J({\bf{x}}_1,t) = \partial_t n({\bf{x}}_1,t) \hspace*{0.5cm} {\text{with}} \hspace*{0.5cm} J({\bf{x}},t) = \left( \frac{1}{2i} \left( \nabla_1
- \nabla_1^{\prime} \right) \rho^{(1)} ({\bf{x}}_1;{\bf{x}}_1^{\prime};t) \right) \Bigr|_{{\bf{x}}_1^{\prime}={\bf{x}}_1}
\label{conteq}
\end{equation}

\noindent
Here $J({\bf{x}}_1,t)$ denotes the particle current density.
The EOM (\ref{rho1eom}) for the single particle density matrix $\rho^{(1)}$ coupled to the integrated
two-particle density matrix $\rho^{(2)}$ can be exploited to obtain an EOM for the natural orbitals and
their occupations \cite{Pernal,Appel,Brics,Rapp,Jansen,Manthe}. This goes as follows. First one inserts the representation (\ref{rho1diag})
of $\rho^{(1)}$ into the EOM (\ref{rho1eom}).
The time-derivative of the natural orbitals is then expanded in the complete set of natural orbitals 
$i \partial_t \phi_k = \alpha_{kk} \phi_k + \sum_{m \ne k} \alpha_{km} \phi_m$  with $\alpha_{kk} \in \mathbb{R}$.
Extracting the global phase ${\cal{P}}_k = exp(-i \int_0^{t} \alpha_{kk} (t^{\prime}) dt^{\prime})$ from the natural orbital $\phi_k$
removes the diagonal term $\propto \alpha_{kk}$. If one subsequently projects onto the natural orbitals 
one arrives at the EOM for the natural populations and the natural orbitals which take on the following appearance

\begin{equation}
 i\partial_t\lambda_n=N(N-1)\int {\rm d}{\bf x}'{\rm d}{\bf x}''{\rm d}{\bf z}
 \,\phi^*_n({\bf x}',t)\,\big[V({\bf x}',{\bf z})-V({\bf x}'',{\bf z}) 
\big]\,\rho_2({\bf x}';{\bf z};{\bf x}'';{\bf z};t)\, \phi_n({\bf x}'',t),
\label{dynnatpop}
\end{equation}

\begin{align}
 i\partial_t\phi_n({\bf x},t)=\sum_{p\neq n}
 \Big(
 &\langle\phi_p|\mathcal{H}_1|\phi_n\rangle+
 \frac{N(N-1)}{\lambda_n-\lambda_p}
 \int {\rm d}{\bf x}'{\rm d}{\bf x}''{\rm d}{\bf z}
 \,\phi^*_p({\bf x}',t)\times \nonumber\\
 &\times\big[V({\bf x}',{\bf z})-V({\bf x}'',{\bf z}) 
\big]\,\rho_2({\bf x}';{\bf z};{\bf x}'';{\bf z};t)\, \phi_n({\bf x}'',t)
 \Big)\phi_p({\bf x},t)
\label{dynnatorb1}
\end{align}

\noindent
Eq.(\ref{dynnatpop}) shows that the rate of change of the natural populations is triggered by interaction with the surrounding particles.
The time dependence of the natural orbitals, however, is influenced by the single particle Hamiltonian
(see first term in eq.(\ref{dynnatorb1})) as well as the interaction. The latter terms are multiplied
by the inverse of the difference of the relevant occupations $(\lambda_n - \lambda_p)^{-1}$ indicating that degeneracies\footnote{In the case 
of a degeneracy, any linear combination of the degenerate natural orbitals is also a natural orbital with the same 
natural occupation number. Thus, the singularity means that the equations of 
motion cannot ``decide'' which linear combination is appropriate for 
a continuous natural orbital trajectory and other criteria have to be employed. 
Whenever we refer to these equations of motion in the following, we tacitly 
assume the absence of degeneracies for simplicity.}
of the occupation numbers lead to corresponding singularities in the expression (\ref{dynnatorb1}). In particular the rate of change
of a given natural orbital is 'driven' by all other natural orbitals as indicated by the sum over all $p \ne n$.

\noindent
Using now the resolution of the identity
for the complete set of natural orbitals $\sum_{p} |\phi_p \rangle \langle \phi_p | = \mathbb{I}$ and extracting the phase
${\cal{L}}_n = exp(+i \int_0^{t} \langle \phi_n | {\cal{H}}_1 | \phi_n \rangle dt^{\prime})$ eliminates the diagonal
single particle term proportional to $\langle \phi_n | {\cal{H}}_1 | \phi_n \rangle$. We therefore finally arrive at
the following EOM for the natural orbitals

\begin{align}
i \partial_t \phi_n ({\bf{x}},t) = {\cal{H}}_1 \phi_n ({\bf{x}},t) + N (N-1) \sum_{p \ne n} (\lambda_n - \lambda_p)^{-1}
\phi_p ({\bf{x}},t) \int \phi_p^{*} ({\bf{x}}^{\prime},t) \left[ V({\bf{x}}^{\prime} , {\bf{z}})
- V({\bf{x}}^{\prime \prime} , {\bf{z}}) \right] \nonumber \\
 \rho^{(2)}({\bf{x}}^{\prime},{\bf{z}};{\bf{x}}^{\prime \prime},{\bf{z}};t)
\phi_n ({\bf{x}}^{\prime \prime},t) d{\bf{x}}^{\prime} d{\bf{x}}^{\prime \prime} d{\bf{z}} 
\label{dynnatorb3}
\end{align}

\noindent
Eq.(\ref{dynnatorb3}) will be the relevant equation below to derive the EOM for symmetry correlators
on the level of individual natural orbitals. From here on we will focus exclusively on a spin-independent formalism, thereby
keeping the same notation as before, for reasons of simplicity. However, we note that the following derivations and results
can straightforwardly be transfered and applied to the spin-dependent case.

\section{Equation of Motion for Symmetry Correlators}

Having established the theoretical background we are now aiming at performing the relevant steps to derive the corresponding
EOM adapted to the presence of local symmetries. First we remind the reader of the fact that for the non-interacting
theory in one spatial dimension it is the constancy of the ITPC (see eqs.(\ref{Q1},\ref{Q2})) that reflect the local symmetry in a given
spatial domain. The ITPC however depend on two symmetry-related spatial positions for a single particle. 
The key idea now for going from a local symmetry formalism for non-interacting particles to interacting many-body systems is to define the 
(canonical) local symmetry correlators as coherences of RDMs between local symmetry related positions. A note of caution is 
appropriate here: we use the term 'correlator' in the sense that a corresponding quantity contains symmetry-related spatial
points which has to be distinguished from the notion of correlations in many-body physics as a beyond mean field correlated motion of particles.
Here, we concentrate on the local symmetry correlators of the one-body RDM. In the following, we will first introduce a local 
symmetry-adapted representation of the equations of motion for the local symmetry correlators or in other words derive
equations of motion for symmetry related coherences of the one-body RDM. Here, 
the crucial step consists in expressing the kinetic part as a divergence of an 
appropriately chosen non-local two-point current density. These equations are 
most transparent in the natural orbital representation of the one-body RDM.
The resulting equation of motion generalizes the ITPC $\tilde{Q}$
(see eq.(\ref{Q2})) to the dynamics of many-body interacting systems, and for this 
reason, we refer to these local symmetry correlators as the canonical ones. In 
a second step we will introduce an anomalous local symmetry correlator
whose equations of motion are derived and we will show that the non-interacting one-dimensional
stationary case of these equations of motion leads to the ITPC $Q$ (see eq.(\ref{Q1})).

\subsection{The Canonical Form of the Symmetry Correlator} \label{CanSymCorr}

We start by rewriting the single particle part of the EOM for $\rho^{(1)}$ (\ref{rho1eom}) in the following way

\begin{equation}
\left( {\cal{H}}_{1} - {\cal{H}}_{1^{\prime}} \right) \rho^{(1)} ({\bf{x}}_1;{\bf{x}}_1^{\prime};t)
= \frac{1}{2}  \left( \nabla^{\prime}_1 + \nabla_1 \right) \left( \nabla^{\prime}_1 - \nabla_1 \right)
\rho^{(1)} ({\bf{x}}_1;{\bf{x}}_1^{\prime};t) + \left( U\left({\bf{x}}_1 \right) - U\left({\bf{x}}_1^{\prime} \right) \right)
\rho^{(1)} ({\bf{x}}_1;{\bf{x}}_1^{\prime};t) 
\label{sppeom}
\end{equation}

\noindent
So far we have assumed that the two spatial arguments ${\bf{x}}_1,{\bf{x}}_1^{\prime}$ are independent. Aiming at the presence
of local symmetries in $U({\bf{x}})$ and knowing about the structure of the ITPC for the non-interacting one-dimensional case
we take ${\bf{x}}_1^{\prime} = {\bf{y}} = \sigma {\bf{x}}_1 + {\bf{L}} = \sigma {\bf{x}} + {\bf{L}}$.
The symmetry mapping ${\bf{y}}={\bf{F}}({\bf{x}})$ corresponds, in general in three dimensions, for 
$\sigma =+1$ to translations by ${\bf{L}}$ and for $\sigma =-1$ to inversions (parity) at the center ${\bf{L}}/2$. 
Now the kinetic term is reformulated as a divergence of an 
appropriately defined non-local current density, which can be achieved in a 
representation-free manner w.r.t.\ the one-body RDM as summarized in Appendix 
\ref{app_basis_free_current}. Much more transparent, however, is the 
reformulation of the kinetic energy on the r.h.s.\ of eq.(\ref{sppeom}) in the 
natural orbital representation:

\begin{eqnarray}
\left( \frac{1}{2N} \right) \sum_{i=1}^{\infty} \lambda_i(t) \nabla_{{\bf{x}}} \left[
\sigma \left( \nabla \phi_i^{*} \right) \left({\bf{y}},t \right) \phi_i \left({\bf{x}},t \right) 
- \phi_i^{*} \left({\bf{y}},t \right) \left( \nabla \phi_i \right) \left({\bf{x}},t \right) \right]
\label{kesppeom}
\end{eqnarray}

\noindent
where $\left( \nabla \phi_i^{*} \right) \left({\bf{y}},t \right) = \nabla_{\bf{x}} \phi_i^{*} ({\bf{x}},t) \Bigr|_{{\bf{x}}={\bf{y}}}$
and the derivative in front of the brackets $\nabla_{{\bf{x}}}$ acts on both arguments ${\bf{x}}$ and ${\bf{y}}={\bf{F}}({\bf{x}})$.
We have therefore succeeded in writing the kinetic energy as a convex sum of divergences of two-point correlator current densities ${\bf j}_i$
belonging to the natural orbitals  $\phi_i$

\begin{equation}
{\bf j}_i \left( {\bf{x}}, {\bf{y}},t \right) =
\left[ \sigma \left( \nabla \phi_i^{*} \right) \left({\bf{y}},t \right) \phi_i \left({\bf{x}},t \right) 
- \phi_i^{*} \left({\bf{y}},t \right) \left( \nabla \phi_i \right) \left({\bf{x}},t \right) \right]
\label{notc}
\end{equation}

\noindent
which is reminescent of the structure of the ITPC obtained in the noninteracting single particle case (see eqs.({\ref{Q1},\ref{Q2})).
In total we therefore arrive at the local symmetry correlator EOM

\begin{equation}
i \partial_t \left[ \sum_{i=1}^{\infty} \lambda_i \left( t \right) 
\left( \phi_i \left({\bf{x}},t \right) \phi_i^{*} \left({\bf{y}},t \right) \right) \right] = 
\frac{1}{2} \nabla_{{\bf{x}}} \sum_{i=1}^{\infty} \lambda_i \left( t \right) {\bf j}_i \left( {\bf{x}}, {\bf{y}},t \right)
\label{toteom} 
\end{equation}
\begin{equation}
 + \left( U\left({\bf{x}} \right) - U\left({\bf{y}} \right) \right) 
\left[ \sum_{i=1}^{\infty} \lambda_i \left( t \right) 
\left( \phi_i \left({\bf{x}},t \right) \phi_i^{*} \left({\bf{y}},t \right) \right) \right]
+ N(N-1) \int \left( V({\bf{x}},{\bf{z}}) - V({\bf{y}},{\bf{z}}) \right) \rho^{(2)} \left({\bf{x}}, {\bf{z}}; {\bf{y}},{\bf{z}};t \right) d{\bf{z}}
\nonumber
\end{equation}

\noindent
where we assumed that ${\bf x}$ lies in the interior of the 
domain $\mathcal{D}$ in order to avoid difficulties regarding the spatial 
derivatives at the boundary of $\mathcal{D}$ and $\bar{\mathcal{D}}$.
Let us pause to think what we have achieved so far by discussing the above continuity-like 
equation (\ref{toteom}) which is one of the central results of the present work. 
The l.h.s. of eq.(\ref{toteom}) represents the time-derivative of the sum of two point symmetry correlators 
$\left( \phi_i \left({\bf{x}},t \right) \phi_i^{*} \left({\bf{y}},t \right) \right)$ for the natural orbitals weighted
by the natural occupation numbers which, as the natural orbitals do, vary in time.
The first term on the r.h.s. is the above-introduced divergence of the sum of natural orbital two-point currents (see eq.(\ref{kesppeom}))
also weighted by the natural populations. The last term on the r.h.s. couples to the second order RDM
$\rho^{(2)} ({\bf{x}},{\bf{z}};{\bf{y}},{\bf{z}};t)$ via the two-body interactions $V$ and 
and will be subject to (functional) approximations in a truncation 
scheme if one aims at closed equations of motion on this order. The collision 
integral may be interpreted as a source term in this continuity-like equation for the local symmetry correlators.
Finally the second term on the r.h.s.
involves the single particle potential $U$. It is exactly this term which disappears domainwise in the
presence of corresponding local translation or inversion symmetries i.e. $U({\bf{x}})=U({\bf{y}})$ $\forall x \in \mathcal{D}$
with ${\bf{y}} = {\bf{F}}({\bf{x}})$. This case, which is of central importance here, leads immediately to the simplified equation
of motion

\begin{equation}
i \partial_t \left[ \sum_{i=1}^{\infty} \lambda_i \left( t \right) 
\left( \phi_i \left({\bf{x}},t \right) \phi_i^{*} \left({\bf{y}},t \right) \right) \right] = 
\frac{1}{2} \nabla_{{\bf{x}}} \sum_{i=1}^{\infty} \lambda_i \left( t \right) {\bf j}_i \left( {\bf{x}}, {\bf{y}},t \right)
\label{toteomsym} 
\end{equation}
\begin{equation}
+ N(N-1) \int \left( V({\bf{x}},{\bf{z}}) - V({\bf{y}},{\bf{z}}) \right) \rho^{(2)} \left({\bf{x}}, {\bf{z}}; {\bf{y}},{\bf{z}};t \right) 
d{\bf{z}}
\nonumber
\end{equation}

\noindent
So for any local inversion or translation symmetry, the dynamics of 
the local symmetry correlator, i.e.\ one-body coherences between local symmetry 
related positions, is not directly driven by the external locally symmetric 
potential $U({\bf x})$ but only, if at all, indirectly via the time-evolution 
of the two-body RDM entering the collision integral.
Let us now discuss the role of the range of the interaction potential by decomposing the collision integral
in eqs.(\ref{toteom},\ref{toteomsym}). With $\bar{\mathcal{D}}=F(\mathcal{D})$, we now define the union
${\mathcal{D}}_T = {\mathcal{D}} \cup {\bar{{\mathcal{D}}}}$ and 
${\mathcal{E}} = {\mathbb{R}}^n \setminus {\mathcal{D}}_T$ as its complement. The collision integral then reads

\begin{equation}
T = T_{\mathcal{D}} + T_{\mathcal{E}} = N(N-1) \lbrace{ \int_{{\mathcal{D}}_T} + \int_{\mathcal{E}} \rbrace} \left( V({\bf{x}},{\bf{z}}) - V({\bf{y}},{\bf{z}}) \right) 
\rho^{(2)} \left({\bf{x}}, {\bf{z}}; {\bf{y}},{\bf{z}};t \right) d{\bf{z}}
\label{collint}
\end{equation}

\noindent
$T_{\mathcal{D}}$ describes the two-body interaction (${\bf{z}} \in {\mathcal{D}}_T$) within the local symmetry domain ${\mathcal{D}}_T$
(note that the case of a gapped local symmetry (see Fig.\ref{fig1}) needs special attention) whereas 
$T_{\mathcal{E}}$ is responsible for the
interaction from outside (${\bf{z}} \notin {\mathcal{D}}_T$). Let us specifically address $T_{\mathcal{D}}$ which can, by using the symmetry 
mapping, be rewritten as follows

\begin{align}
T_{\mathcal{D}}  = N(N-1) \left( \int_{{\mathcal{D}}} \left( V({\bf{x}},{\bf{z}}) - V({\bf{y}},{\bf{z}}) \right) 
\rho^{(2)} \left({\bf{x}}, {\bf{z}}; {\bf{y}},{\bf{z}};t \right) d{\bf{z}} \right)  \nonumber \\
+ N(N-1) \left( \int_{{\mathcal{D}}} \left( V({\bf{x}},{\bf{F}}({\bf{z}}^{\prime})) - V({\bf{y}},{\bf{F}}({\bf{z}}^{\prime})) \right) 
\rho^{(2)} \left({\bf{x}}, {\bf{F}}({\bf{z}}^{\prime}); {\bf{y}},{\bf{F}}({\bf{z}}^{\prime});t \right) d{\bf{z}}^{\prime} \right)
\label{collintD}
\end{align}

\noindent
If the interaction potential depends only on the distance between the particles i.e. $V({\bf{x}},{\bf{z}}) = V(|{\bf{x}}-{\bf{z}}|)$
we have $V(|{\bf{x}}-{\bf{z}}|) = V(|{\bf{y}}-{\bf{F}}({\bf{z}})|)$ and obtain 

\begin{align}
T_{\mathcal{D}}  = N(N-1) \int_{{\mathcal{D}}} V(|{\bf{x}}-{\bf{z}}|) \left(
\rho^{(2)} \left({\bf{x}}, {\bf{z}}; {\bf{y}},{\bf{z}};t \right)
- \rho^{(2)} \left({\bf{x}}, {\bf{F}}({\bf{z}}); {\bf{y}},{\bf{F}}({\bf{z}});t \right) \right) d{\bf{z}}  \nonumber \\ 
+ N(N-1) \int_{{\mathcal{D}}} \left( V(|{\bf{x}}-{\bf{F}}({\bf{z}})|) 
\rho^{(2)} \left({\bf{x}}, {\bf{F}}({\bf{z}}); {\bf{y}},{\bf{F}}({\bf{z}});t \right)
- V(|{\bf{y}}-{\bf{z}}|) \rho^{(2)} \left({\bf{x}}, {\bf{z}}; {\bf{y}},{\bf{z}};t \right) \right) d{\bf{z}}
\label{collintD2}
\end{align}

\noindent
In the above expression the first integral contains the intradomain particle interactions and the second one
describes the particle interactions between the domains ${\mathcal{D}}$ and ${\bar{{\mathcal{D}}}}$. This second term provides only a contribution
if the range of the interaction exceeds the minimal distance between ${\mathcal{D}}$ and ${\bar{{\mathcal{D}}}}$. A special case is the local
inversion symmetry in which case we have $V(|{\bf{x}}-{\bf{F}}({\bf{z}})|) = V(|{\bf{F}}({\bf{x}})-{\bf{z}}|)$. Then the
contribution $T_{\mathcal{D}}$ reads

\begin{align}
T_{\mathcal{D}}  = N(N-1) \int_{{\mathcal{D}}} \left(V(|{\bf{x}}-{\bf{z}}|) - V(|{\bf{y}}-{\bf{z}}|) \right)  \left(
\rho^{(2)} \left({\bf{x}}, {\bf{z}}; {\bf{y}},{\bf{z}};t \right) 
- \rho^{(2)} \left({\bf{x}}, {\bf{F}}({\bf{z}}); {\bf{y}},{\bf{F}}({\bf{z}});t \right) \right) d{\bf{z}}
\label{collintD3}
\end{align}

\noindent
In ultracold quantum gases \cite{Pethick} the interaction in the course of the ultracold collisions can be modeled by a so-called contact potential of the form
$V({\bf{x}},{\bf{x}}^{\prime}) = g \delta({\bf{x}}-{\bf{x}}^{\prime})$ which is the extreme limit of a short-range interaction.
In this case $T_\mathcal{E}$ vanishes and the collision integral in total takes the following appearance

\begin{align}
T  = N(N-1) g \left( \rho^{(2)} \left({\bf{x}}, {\bf{x}}; {\bf{y}},{\bf{x}};t \right) 
- \rho^{(2)} \left({\bf{x}}, {\bf{y}}; {\bf{y}},{\bf{y}};t \right) \right) 
\label{collintD4}
\end{align}

\noindent
This concludes our discussion of the collision integral $T$ and we turn now back to our equations
of motion for the symmetry correlators.
Eq.(\ref{toteom}) possesses an integral representation which can be obtained by integrating it over a volume $V$ and employing
Gauss (divergence) theorem which yields finally

\begin{equation}
i \partial_t \int_{V} \left[ \sum_{i=1}^{\infty} \lambda_i \left( t \right) 
\left( \phi_i \left({\bf{x}},t \right) \phi_i^{*} \left({\bf{y}},t \right) \right) \right] d{\bf x} = 
\frac{1}{2} \oint_{S(V)} \left( \sum_{i=1}^{\infty} \lambda_i \left( t \right) {\bf j}_i \left( {\bf{x}}, {\bf{y}},t \right) \right) d{\bf S}
\label{intreprtoteom} 
\end{equation}
\begin{equation}
 + \int_{V} \left( U\left({\bf{x}} \right) - U\left({\bf{y}} \right) \right) 
\left[ \sum_{i=1}^{\infty} \lambda_i \left( t \right) 
\left( \phi_i \left({\bf{x}},t \right) \phi_i^{*} \left({\bf{y}},t \right) \right) \right] d{\bf x}
\nonumber
\end{equation}
\begin{equation}
+ N(N-1) \int_{V} \int \left( V({\bf{x}},{\bf{z}}) - V({\bf{y}},{\bf{z}}) \right) \rho^{(2)} \left({\bf{x}}, {\bf{z}}; {\bf{y}},{\bf{z}};t \right) d{\bf{z}}
d {\bf x}
\nonumber
\end{equation}

\noindent
where $S(V)$ is the surface of the volume $V$ and $d{\bf S}$ denotes the normal vector of the surface element\footnote{A word of caution 
is in order here: Eq.(\ref{intreprtoteom}) is valid if $V$ lies in the interior of $D$. Otherwise, one has to extend the local 
coordinate transformation ${\bf{F}}({\bf x})$ to a global one and carefully inspect whether the kinetic term
results in possibly even distribution valued singularities on the surface of $\mathcal{D}$.}.
The time-dependent symmetry correlator change integrated over a volume $V$ involves therefore the flux out of this volume
due to the convex sum of the weighted currents ${\bf j}_i$. Both, the external single particle potential and the
interaction occur in their equally volume-integrated form. Again, in the presence of corresponding local symmetries
the term due to the external potential vanishes per construction.

\noindent
In the above EOM (\ref{toteom},\ref{toteomsym}) the symmetry correlators appear in the form of convex
sums, i.e. for both the time-derivative as well as the term involving the currents they are weighted by the natural
populations and summed over all natural orbitals. It is however possible to derive EOM for the symmetry
correlators for individual natural orbitals. Our starting-point is here eq.(\ref{dynnatorb3}). Constructing the quantity
$(i \partial_t \phi_n^{*} ({\bf{y}},t)) \phi_n ({\bf{x}},t) + (i \partial_t \phi_n ({\bf{x}},t)) \phi_n^{*} ({\bf{y}},t)
= i \partial_t (\phi_n ({\bf{x}},t) \phi_n^{*} ({\bf{y}},t))$ where again ${\bf{x}},{\bf{y}}$ are symmetry related via
${\bf{y}}= F({\bf{x}})$ and reformulating the action of the Laplacian as a total divergence as well as exploiting 
the hermiticity of the density matrix we arrive at the desired EOM for a single natural orbital-based symmetry correlator

\begin{align}
i \partial_t \left( \phi_n ({\bf{x}},t) \phi_n^{*} ({\bf{y}},t) \right) = \frac{1}{2} \nabla_{\bf{x}} {\bf{j}}_n ({\bf{x}},{\bf{y}},t)
+ \left( U({\bf{x}}) - U({\bf{y}}) \right) \phi_n ({\bf{x}},t) \phi_n^{*} ({\bf{y}},t) \nonumber \\ + N(N-1)
\sum_{p=1, p \ne n}^{\infty} \left( \lambda_n - \lambda_p \right)^{-1} \left( I_{pn} \phi_p ({\bf{x}},t) \phi_n^{*} ({\bf{y}},t)
+ I_{np} \phi_n ({\bf{x}},t) \phi_p^{*} ({\bf{y}},t) \right)
\label{onenatsymeom}
\end{align}

\noindent
where ${\bf{j}}_n ({\bf{x}},{\bf{y}},t)$ is the single natural orbital symmetry correlator current (see eq.(\ref{notc}))
and $I_{pn}$ denote the matrix elements of the collision integral in the natural orbital basis

\begin{align}
I_{pn} = \int d{\bf{x}}^{\prime} d{\bf{x}}^{\prime \prime} d{\bf{z}}~~\phi_n ({\bf{x}}^{\prime \prime},t) \phi_p^{*} ({\bf{x}}^{\prime},t) 
 \left[ V({\bf{x}}^{\prime},{\bf{z}}) - V({\bf{x}}^{\prime \prime},{\bf{z}}) \right]
\rho^{(2)} ({\bf{x}}^{\prime},{\bf{z}};{\bf{x}}^{\prime \prime},{\bf{z}};t)
\label{IPN}
\end{align}

\noindent
Local symmetry implies again $U({\bf{x}}) = U({\bf{y}})$ and one should note that in this case the
EOM for the individual symmetry correlators 
depend only indirectly on the external potential, namely via $\rho^{(2)}$ as we have already
observed above for the total local symmetry correlator. Eq.(\ref{onenatsymeom}) shows that the dynamics of the symmetry
correlator of a single natural orbital involves apart from the corresponding correlator current ${\bf{j}}_n$ in particular
all off-diagonal correlators $\phi_n ({\bf{x}}^{\prime \prime},t) \phi_p^{*} ({\bf{x}}^{\prime},t)$
involving the other natural orbitals. We remark that a corresponding EOM can be derived for the dynamics
of these off-diagonal correlators as done above for the diagonal term.
In passing, we note that by considering the trivial map ${\bf y}=F({\bf 
x})={\bf x}$, eq.(\ref{onenatsymeom}) turns into a continuity equation with 
a local source term for the individual natural orbital densities. This source 
term is induced by the collision integral, which couples the natural orbital 
densities to the other natural orbitals, and vanishes when being integrated over 
the whole space, which reflects the conservation of probability.

\noindent
In the following subsections we will derive and discuss several special cases for the above equation. This includes first the
non-interacting case where we derive a continuity equation for the two-point symmetry correlators and recover the case of the
ITPC for a time-independent theory. The two subsequent subsections discuss the mean field theory for bosons and fermions, respectively.

\subsubsection{Non-interacting Particle Systems}

Let us first focus on the case of non-interacting particles $V=0$ in the presence of local symmetries $U({\bf{x}})=U({\bf{y}})$ which
leads via eq.(\ref{toteomsym}) to

\begin{equation}
i \partial_t \left[ \sum_{i=1}^{\infty} \lambda_i \left( t \right)
\left( \phi_i \left({\bf{x}},t \right) \phi_i^{*} \left({\bf{y}},t \right) \right) \right] =
\frac{1}{2} \nabla_{{\bf{x}}} \sum_{i=1}^{\infty} \lambda_i \left( t \right) {\bf j}_i \left( {\bf{x}}, {\bf{y}},t \right)
\label{contineq}
\end{equation}

\noindent
and via eq.(\ref{onenatsymeom}) to

\begin{equation}
i \partial_t \left( \phi_i \left({\bf{x}},t \right) \phi_i^{*} \left({\bf{y}},t \right) \right) =
\frac{1}{2} \nabla_{{\bf{x}}} {\bf j}_i \left( {\bf{x}}, {\bf{y}},t \right)
\label{contineq2}
\end{equation}

\noindent
This is a generalized continuity equation for the two-point symmetry correlators. Reducing to a single particle in 
one dimension which is the case previously treated extensively 
\cite{Kalozoumis2013a,Kalozoumis2014a,Kalozoumis2014b,Kalozoumis2013b,Kalozoumis2015a,Morfonios2014,
Kalozoumis2015b,Zambetakis2016,Kalozoumis2016a,Wulf2016}
we obtain

\begin{equation}
i \partial_t \left[ \phi \left(x,t \right) \phi^{*} \left(y,t \right) \right] =
\frac{1}{2} \partial_{x} j \left( x, y,t \right)
\label{onepartconteom}
\end{equation}

\noindent
with $ j \left( x, y,t \right) = \left[ \sigma \left( \partial_x \phi^{*} \right) \left(y,t \right)
\phi \left(x,t \right) - \phi^{*} \left(y,t \right) \left( \partial_x \phi \right) \left(x,t \right) \right]$
where one should always keep in mind that $y=F(x)$ and the derivative acts therefore on both $x$ and $y$.
Focusing on the stationary i.e. time-independent Schr\"odinger (or Helmholtz) equation $\phi (x,t) = exp(-i E t) \phi(x)$
with the single particle energy eigenvalue $E$ and the stationary single particle bound or scattering state $\phi(x)$ the time
derivative vanishes due to the time independence of the product $\phi \left(x,t \right) \phi^{*} \left(y,t \right)$.
Integrating the r.h.s. of eq.(\ref{onepartconteom}) yields then

\begin{equation}
\frac{1}{2i} j \left( x, y,t \right) = \frac{1}{2i} \left[ \sigma \left( \partial_x \phi^{*} \right) \left(y,t \right)
\phi \left(x,t \right) - \phi^{*} \left(y,t \right) \left( \partial_x \phi \right) \left(x,t \right) \right] = C
\label{Qtildeconst}
\end{equation}

\noindent
which is identical to the complex conjugate of the previously discovered \cite{Kalozoumis2013b,Kalozoumis2014a} invariant two-point current 
correlator $\widetilde{Q}$ given in eq.(\ref{Q2}). The latter is constant in each corresponding domain of local symmetry.
Our general theoretical framework therefore contains the already known special case of stationary states in one dimension.
This remark refers to the ITPC $\widetilde{Q}$. The case of the ITPC $Q$ will be treated separately below since
it goes beyond the canonial ket bra structure of the density matrix formalism.

\subsubsection{Mean-field theory: Bosons}

Let us focus next on the case of a bosonic many-body ensemble where we have in particular in mind
ultracold quantum gases which can in many situations successfully be described by a mean field Gross-Pitaevskii 
equation that assumes the particles to be condensed in a single orbital \cite{Pitaevskii,Pethick}
i.e.\ that there is a natural orbital with $\lambda_i=N$. 
The many-body wave function then reads $\Psi({\bf{x}}_1,...,{\bf{x}}_N,t) = \phi({\bf{x}}_1,t) \cdot .... \cdot \phi({\bf{x}}_N,t)$
which trivially obeys the particle exchange symmetry. The resulting single particle and two particle density matrices
read $\rho^{(1)} ({\bf{x}}_1;{\bf{x}}_1^{\prime};t) = \phi({\bf{x}}_1,t) \phi^{*} ({\bf{x}}_1^{\prime},t)$ and
$\rho^{(2)} ({\bf{x}}_1,{\bf{x}}_2;{\bf{x}}_1^{\prime},{\bf{x}}_2^{\prime};t) = \phi({\bf{x}}_1,t) \phi({\bf{x}}_2,t) 
\phi^{*} ({\bf{x}}_1^{\prime},t) \phi^{*} ({\bf{x}}_2^{\prime},t)$ respectively.
The interaction among the bosons can in the ultracold regime be described by a so-called contact potential
$V({\bf{x}}_i - {\bf{x}}_j) = g \delta ({\bf{x}}_i - {\bf{x}}_j)$ which depends on a single parameter, the $s$-wave scattering length
contained in $g$. Assuming again a locally symmetric potential $U({\bf{x}})=U({\bf{y}})$ the above EOM (\ref{toteomsym}) 
for the symmetry correlator simplifies to 

\begin{eqnarray}
i \partial_t \left( \phi \left({\bf{x}},t \right) \phi^{*} \left({\bf{y}},t \right) \right) =
\frac{1}{2} \nabla_{{\bf{x}}} {\bf j}({\bf{x}}, {\bf{y}}, t) + 
(N-1)g \left[|\phi({\bf{x}},t)|^2 - |\phi({\bf{y}},t)|^2 \right] \phi({\bf{x}},t) \phi^{*}({\bf{y}},t) 
\label{boson1}
\end{eqnarray}

\noindent
with ${\bf j}$ given in eq.(\ref{notc}) for a single natural orbital.
Now, due to the interactions on the mean field level a 'source term' of a nonlinear character emerges. 
This term is proportional to the difference $\left[|\phi({\bf{x}},t)|^2 - |\phi({\bf{y}},t)|^2 \right]$
which shows the remarkable fact that if the magnitude of the wave function follows the local symmetry $|\phi({\bf{x}},t)| = |\phi({\bf{y}},t)|$ 
then the symmetry correlator behaves as if there would be no interactions present on the mean field level.
This is of particular importance for the case of a stationary state for which eq.(\ref{boson1}) leads to

\begin{eqnarray}
\frac{1}{2} \nabla_{{\bf{x}}} {\bf j}({\bf{x}}, {\bf{y}}) =
(N-1)g \left[|\phi({\bf{y}})|^2 - |\phi({\bf{x}})|^2 \right] \phi({\bf{x}}) \phi^{*}({\bf{y}})
\label{statboson}
\end{eqnarray}

\noindent
This is reminescent of the occurence of the so-called symmetric perfectly transmitting resonances
\cite{Kalozoumis2013b} which indeed possess the property that the magnitude of the wave function follows
the local symmetries of the underlying potential $U$. 

\subsubsection{Mean-field theory: Fermions}

Within the mean field theory of fermions, namely Hartree-Fock theory, $N$ electrons occupy $N$ orbitals forming
a Slater determinant that is completely antisymmetric with respect to particle exchange (as indicated previously,
the spin degrees of freedom are ignored tacitly here). We have therefore $N$ natural orbitals with occupation
numbers equal to one. Exploiting once again local symmetry $U({\bf{x}}) = U({\bf{y}})$ of the external
potential the corresponding EOM reads

\begin{equation}
i \partial_t \left( \sum_{i=1}^{N} \phi_i \left({\bf{x}},t \right) \phi_i^{*} \left({\bf{y}},t \right) \right) = 
\frac{1}{2} \nabla_{{\bf{x}}} \sum_{i=1}^{N} {\bf j}_i \left( {\bf{x}}, {\bf{y}},t \right)
\label{HFeom} 
\end{equation}
\begin{equation}
+ \int \left( V({\bf{x}},{\bf{z}}) - V({\bf{y}},{\bf{z}}) \right) 
\sum_{i,j=1}^{N} \left[ 
\phi_i({\bf{x}},t) \phi_i^{*}({\bf{y}},t) |\phi_j({\bf{z}},t)|^2
- \phi_j({\bf{x}},t) \phi_i^{*}({\bf{y}},t) \phi_i({\bf{z}},t) \phi_j^{*}({\bf{z}},t) \right]
d{\bf{z}}
\nonumber
\end{equation}

\noindent
where now due to the Pauli principle $N$ natural orbital symmetry correlator currents contribute.
The interaction term contains the corresponding Coulomb and exchange terms, respectively. The stationary
case follows immediately as

\begin{equation}
\frac{1}{2} \nabla_{{\bf{x}}} \sum_{i=1}^{N} {\bf j}_i \left( {\bf{x}}, {\bf{y}} \right) =
 \int \left( V({\bf{y}},{\bf{z}}) - V({\bf{x}},{\bf{z}}) \right) 
\sum_{i,j=1}^{N} \left[ \phi_i({\bf{x}}) \phi_i^{*}({\bf{y}}) |\phi_j({\bf{z}})|^2
- \phi_j({\bf{x}}) \phi_i^{*}({\bf{y}}) \phi_i({\bf{z}}) \phi_j^{*}({\bf{z}}) \right]
d{\bf{z}}
\nonumber
\end{equation}

\noindent
We do not address the problems which arise due to the mean field approximation in the equations of
motion but refer for this to the literature (see refs. \cite{Bonitz,Akbari} and references therein).

\subsection{The Anomalous Symmetry Correlator}

The theoretical framework presented so far is derived from the BBGKY-hierarchy which has the canonical
ket bra structure such that the scalar product of Hilbert space can be exploited. The stationary state
case leads then to the canceling of the dynamical phase factor and single (natural) orbital orthogonality within
e.g. mean field theory can be exploited. As a matter of fact, however, the ITPC
$Q$ in eq.(\ref{Q1}) is not of this canonical structure but is still a relevant quantity as has been
shown via the generalization of the parity and Bloch theorem \cite{Kalozoumis2014a} or in applications to parity-time
symmetry breaking of scattering states \cite{Kalozoumis2014b} where $Q$ plays a decisive role to construct 
the underlying phase diagram. So, in what follows, we 'leave safe grounds' based on the canonical structure
and establish the anomalous EOM of the generalization of $Q$ to higher dimensional many-body interacting
systems. Though this might seem disturbing at first glance, it is indeed motivated by the above-indicated
importance of the ITPC $Q$.

Let us start by defining the quantity 

\begin{equation}
\gamma^{(N)} ({\bf{x}}_1,...,{\bf{x}}_N;{\bf{x}}_1^{\prime},...,{\bf{x}}_N^{\prime};t)
= \Psi({\bf{x}}_1,...,{\bf{x}}_N,t) \Psi({\bf{x}}_1^{\prime},...,{\bf{x}}_N^{\prime},t)
\label{gam1}
\end{equation}

\noindent
which is a complex symmetric quantity as compared to the hermitian density matrix $\rho^{(N)}$. We can then construct
the corresponding reduced quantities $\gamma^{(n)}$ of order $n$

\begin{equation}
\gamma^{(n)} (X_n;X_n^{\prime};t) = \int \Psi (X_n^{\prime},X_n^c,t) \Psi (X_n,X_n^c,t) dX_n^c
\label{gam2}
\end{equation}

\noindent
In order to derive an EOM for $\gamma^{(n)}$ we define the time-derivative \footnote{This time-derivative
is reminiscent of a multiple-time calculus $i \hat{\partial}_t \gamma^{(N)} =  \left(i \partial_t  - i \partial_{t^{\prime}} \right) \Psi(X_N,t) 
\Psi (X_N^{\prime},t^{\prime}) \vert_{t^{\prime} = t}$}.

\begin{equation}
i \hat{\partial}_t \gamma^{(N)} = i  \left( \partial_t  \Psi(X_N,t) \right) \Psi (X_N^{\prime},t)
- i  \left( \partial_t  \Psi(X_N^{\prime},t) \right) \Psi (X_N,t)
\label{gam3}
\end{equation}

\noindent
and correspondingly for $\gamma^{(n)}$ where the time derivative acts under the spatial integral.
Following the same scheme as in the derivation of the BBGKY hierarchy, i.e. the steps which have been
outlined above (see section \ref{BBGKYsec}), we can again partition the Hamiltonian in the equation of
motion for $\gamma^{(n)}$ into degrees of freedom of $\gamma^{(n)}$ and those which are traced out,
i.e. integrated over. Partial integration is now sufficient to get rid of the terms where the corresponding Hamiltonian acts
on the integrated spatial variables and, together with permutation symmetry, we obtain the EOM for $\gamma^{(n)}$

\begin{eqnarray}
i \hat{\partial}_t \gamma^{(n)} (X_n;X_n^{\prime};t) =
\left( {\cal{H}}_{1...n} - {\cal{H}}_{1^{\prime}...n^{\prime}} \right) \gamma^{(n)} (X_n;X_n^{\prime};t) \\ \nonumber
+ (N-n) \sum_{i=1}^{n} \int \left( V({\bf{x}}_i , {\bf{x}}_{n+1}) - V({\bf{x}}_i^{\prime} , {\bf{x}}_{n+1})  \right)
\gamma^{(n+1)} (X_n,{\bf{x}}_{n+1};X_n^{\prime},{\bf{x}}_{n+1};t) d{\bf{x}}_{n+1}
\label{gam4}
\end{eqnarray}

\noindent
and for the lowest order equation for $\gamma^{(1)}$

\begin{eqnarray}
i \hat{\partial}_t \gamma^{(1)} ({\bf{x}}_1;{\bf{x}}_1^{\prime};t) =
\left( {\cal{H}}_{1} - {\cal{H}}_{1^{\prime}} \right) \gamma^{(1)} ({\bf{x}}_1;{\bf{x}}_1^{\prime};t) \\ \nonumber
+ (N-1) \int \left( V({\bf{x}}_1 , {\bf{x}}_{2}) - V({\bf{x}}_1^{\prime} , {\bf{x}}_{2})  \right)
\gamma^{(2)} ({\bf{x}}_1,{\bf{x}}_{2};{\bf{x}}_1^{\prime},{\bf{x}}_{2};t) d{\bf{x}}_{2}
\label{gam5}
\end{eqnarray}

\noindent
Due to the fact that $\gamma^{(n)}$ is complex symmetric (and not hermitian) it is not necessarily diagonalizable.
However, we do assume here that it can be diagonalized in which case it has an eigendecomposition which reflects the
complex symmetry \cite{Horn}. More precisely, diagonalization is possible if and only if the corresponding eigenvector
matrix $Z$ can be chosen such that $Z^T \gamma Z = D$ and $Z^T Z = 1$, where $Z$ is complex orthogonal and $D$ is the
diagonal eigenvalue matrix. The complex orthogonality of $Z$ reflects the complex symmetry of $\gamma$.
This leads us to the representation

\begin{equation}
\gamma^{(1)} ({\bf{x}}_1;{\bf{x^{\prime}}}_1;t) = \sum_{i} \mu_i(t) \chi_i({\bf{x}}_1;t) \chi_i({\bf{x}}^{\prime}_1;t)
\label{gamma1diag}
\end{equation}

\noindent
It is important to note here that, due to the complex symmetry, the orbitals $\chi_i$ do not constitute an orthonormal
basis in the single particle Hilberg space but remain a basis under the above specified assumption. 

Following the same line of argumentation as in section \ref{CanSymCorr}, i.e. the introduction of the two symmetry (inversion, translation)
related arguments ${\bf{x}}$ and ${\bf{y}}$ instead of the independent variables ${\bf{x}}_1,{\bf{x^{\prime}}}_1$ and the
formulation of the kinetic energy terms as a divergence of currents, leads to the following result

\begin{equation}
i \hat{\partial}_t \left[ \sum_{i=1}^{\infty} \mu_i \left( t \right) 
\left( \chi_i \left({\bf{x}},t \right) \chi_i \left({\bf{y}},t \right) \right) \right] = 
\frac{1}{2} \nabla_{{\bf{x}}} \sum_{i=1}^{\infty} \mu_i \left( t \right) {\bf j}^{a}_i \left( {\bf{x}}, {\bf{y}},t \right)
\label{toteomanom} 
\end{equation}

\begin{equation}
 + \left( U\left({\bf{x}} \right) - U\left({\bf{y}} \right) \right) 
\left[ \sum_{i=1}^{\infty} \mu_i \left( t \right) 
\left( \chi_i \left({\bf{x}},t \right) \chi_i \left({\bf{y}},t \right) \right) \right]
+ (N-1) \int \left( V({\bf{x}},{\bf{z}}) - V({\bf{y}},{\bf{z}}) \right) \gamma^{(2)} \left({\bf{x}}, {\bf{z}}; {\bf{y}},{\bf{z}};t \right) d{\bf{z}}
\nonumber
\end{equation}

\noindent
which is the EOM for the anomalous symmetry correlator and contains the corresponding anomalous
correlator current

\begin{equation}
{\bf j}^{a}_i \left( {\bf{x}}, {\bf{y}},t \right) =
\left[ \sigma \left( \nabla \chi_i \right) \left({\bf{y}},t \right) \chi_i \left({\bf{x}},t \right) 
- \chi_i \left({\bf{y}},t \right) \left( \nabla \chi_i \right) \left({\bf{x}},t \right) \right]
\label{anomc}
\end{equation}

\noindent
In the presence of local symmetry $U({\bf{x}})=U({\bf{y}})$ and the corresponding above potential term vanishes, in
complete analogy to the case of the canonical current in section \ref{CanSymCorr}. Addressing the non-interacting
case we arrive at an anomalous correlator continuity equation

\begin{equation}
\hat{\partial}_t \left[ \sum_{i=1}^{\infty} \mu_i \left( t \right) 
\left( \chi_i \left({\bf{x}},t \right) \chi_i \left({\bf{y}},t \right) \right) \right] = 
\frac{1}{2i} \nabla_{{\bf{x}}} \sum_{i=1}^{\infty} \mu_i \left( t \right) {\bf j}^{a}_i \left( {\bf{x}}, {\bf{y}},t \right)
\label{contcoranom} 
\end{equation}

\noindent
Further specializing to the case of a stationary eigenstate of a single particle with its dynamical phase factor $exp(iEt)$
and energy eigenvalue $E$ we obtain

\begin{equation}
\frac{1}{2i} \nabla_{{\bf{x}}} \left( t \right) {\bf j}^{a}_i \left( {\bf{x}}, {\bf{y}},t \right) = 0
\label{spc}
\end{equation}

\noindent
which can be directly integrated for the one-dimensional case and yields the constancy of $Q$ in eq.(\ref{Q1}).
Therefore, the well-known invariant correlator current $Q$ is recovered for this special case of local symmetries
within the present framework of EOM. Addressing the situation of the presence of interactions
is not as transparent as in the case of the canonical symmetry correlator here.
This is due to the fact that the anomalous structure of $\gamma^{(n)}$ does not allow to exploit the underlying
scalar product of Hilbert space when tracing over the corresponding subspace as done e.g. by representing the
wave function in terms of permanents (bosons) or determinants (fermions).

\noindent
Finally, we emphasize that it is not unique how to construct an 
anomalous (one-body) local symmetry correlator for an interacting many-body 
ensemble such that its stationary non-interacting limit leads to the constancy 
of the ITPC $Q$ in one-dimension. In Appendix \ref{Qadds}, we discuss a distinct, 
alternative approach based on the equations of motion for the natural orbitals. 
Further experience resulting in additional requirements on the construction of 
the anomalous symmetry correlators is needed in order to decide which of the 
discussed (or other conceivable) constructions is most appropriate.

\section{Conclusions}

Local order and symmetries are very widespread in complex and composite systems. They can occur for ground or
metastable states in a diversity of physical setups and, beyond their natural occurence, local symmetries might
be a tool to design certain properties of wave mechanical systems such as perfect transmission or localization.
Still, very little is known about the theoretical framework which describes e.g. quantum systems with local
symmetries. It is only recently that it has been shown \cite{Kalozoumis2013a,Kalozoumis2014a,Kalozoumis2014b,
Kalozoumis2013b,Kalozoumis2015a,Morfonios2014,Kalozoumis2015b,Zambetakis2016,Kalozoumis2016a,Wulf2016}
that the unique feature of the presence of local symmetries are the existence of invariant two-point correlator
currents. However, all of these works address the case of one-dimensional stationary and non-interacting wave mechanical
setups ranging from acoustics, optics to quantum mechanics and make no statement how a more general
concept of local symmetries could look like, in particular in the case of dynamically interacting quantum systems.
The purpose of the present work has been to fill this gap and establish a theoretical framework and corresponding
equations of motion for higher dimensional and interacting many-body quantum systems. We hereby focus on local
discrete translation and inversion symmetries. 

While the straightforward extension of the calculus providing
the invariant two-point correlator currents in the noninteracting one-dimensional case to many-body systems
is plagued by problems, such as the long-range character of the interaction versus the locality of the symmetry,
a completely different approach turns out to be successful. Instead of addressing
all microscopic degrees of freedom via the many-body Schr\"odinger equation directly, integrating out most of the
degrees of freedom and remaining with only a few effective ones reveals to be a very fruitful pathway. Specifically
we have employed the framework of the BBGKY hierarchy
which allows to include both the external potential and the interaction among the particles. The canonical ket bra
structure of this hierarchy of equations of motions for the density matrices leads then to an
equation of motion for the first order symmetry correlators that contains a convex sum (incoherent superposition) of generalized two-point currents 
based on the natural orbitals. Employing the equations of motion for the individual natural orbitals
the equation governing the temporal evolution of the individual natural orbital correlators could be established.
We have analyzed several special cases, such as the bosonic and fermionic mean field approximation, for 
which this equation of motion is closed since the second order density matrix is known explicitly. Our approach
is the desired generalization of the two-point invariant currents $\widetilde{Q}$ of the stationary one-dimensional and non-interacting
theory with local symmetries to the dynamics of interacting particles in arbitrary dimensions as we have shown
explicitly. In a second step we have developed an equation of motion for the corresponding anomalous correlator specializing to the
invariant two-point current $Q$ which is responsible for the symmetry breaking. 

The present work lays down the theoretical framework for the dynamics of interacting particles in the presence
of local symmetries. Let us mention some immediate questions that could be followed up. Within the BBGKY hierarchy 
it would be natural to try to establish similar equations of motion for symmetry correlators of higher than
first order. We did not pursue this pathway here since our focus was on the generalization of the theoretical framework
of the invariant two-point currents $Q,\widetilde{Q}$ which is, as shown, contained in the first order equations.
Another aspect that might be worth while considered, is that the present approach could be even helpful in
the case of the absence (or approximate presence) of local symmetries. Then the source term proportional to 
$(U({\bf{x}})-U({\bf{y}}))$ 'drives' the continuity equation of the symmetry correlator as the interaction term does.
Variation of the parameter ${\bf{L}}$ of the symmetry transform might complete the partial information available by the
symmetry correlator (note that the canonical symmetry correlator provides a subspace information of the complete density matrix).
Beyond the above other transforms than translation and inversion could be considered and might lead, in certain 
cases, to helpful further relations. Finally we remark that the equation of motion of the anomalous symmetry
correlator certainly needs further inspection. It was here established to demonstrate that a generalization of 
the symmetry breaking parameter $Q$ indeed exists to the case of the dynamics of higher dimensional
interacting particle systems.
\vspace*{0.7cm}

\noindent
{\bf{Acknowledgment}}\\
We thank J. Schirmer for a careful reading of the manuscript.

\appendix

\section{Representation-free EOM for local symmetry correlator}
\label{app_basis_free_current}

In order to derive an equation of motion for the local symmetry correlator 
which does not rely on the natural orbital representation of the one-body RDM, 
we write the kinetic term of eq.(\ref{rho1eom}) with ${\bf x}_1={\bf 
x}$ and ${\bf x}_1'=F({\bf x})$ as:
\begin{equation} 
\frac{1}{2}(\Delta'-\Delta) \rho_1( { \bf x};{\bf 
x}';t)\Big|_{{\bf x}'=F({\bf x})}= \frac{1}{2}
(\sigma\nabla'+\nabla)(\sigma\nabla'-\nabla)\rho_1( { \bf x};{\bf 
x}';t)\Big|_{{\bf x}'=F({\bf x})}
\end{equation}
Defining ${\bf d}({\bf x},{\bf x}')\equiv (\sigma\nabla'-\nabla)\rho_1( { \bf 
x};{\bf x}';t)$, one can convince oneself that $\nabla {\bf d}({\bf x},F({\bf 
x}))= (\sigma\nabla'+\nabla){\bf d}({\bf x},{\bf x}')$ with 
${\bf x}'$ taken to be $F({\bf x})$. Finally, one may express $ {\bf d}({\bf 
x},F({\bf x}))$ in terms of the abstract momentum operator $\hat {\bf p}$ and 
the one-body reduced density operator $\hat \rho_1$ (note the atomic units):
\begin{equation}
 {\bf d}({\bf x},F({\bf x})) = \frac{1}{i}\langle {\bf x}|\,\big[\,\hat {\bf 
p}\hat\rho_1+\sigma\hat\rho_1\hat {\bf p}\, \big]\,|F({\bf x})\rangle
\end{equation}
where $|{\bf x}\rangle$, $|F({\bf x})\rangle$ denote the position operator 
eigenvectors with the corresponding eigenvalues ${\bf x}$, $F({\bf x})$, 
respectively. In this way, we arrive at the representation-free equation of 
motion for the local symmetry correlator by replacing the first term on the 
r.h.s.\ of eq.(\ref{toteom}) by
\begin{equation}
 \frac{1}{2i} \nabla_{\bf x} \,\langle {\bf x}|\,\big[\,\hat {\bf 
p}\hat\rho_1+\sigma\hat\rho_1\hat {\bf p}\, \big]\,|F({\bf x})\rangle
\end{equation}
where the derivative also acts on the label of the bra vector, $F({\bf x})$. 
Thereby, we find that the non-local current density associated with the 
local symmetry correlator essentially equals the local symmetry related 
off-diagonal matrix elements of the (anti-) commutator between the momentum and 
the one-body reduced density operator for the case of a local (translation) 
parity symmetry.

\section{An alternative construction of the anomalous local symmetry correlator}
\label{Qadds}

As an alternative to the non-canonical structure of the $\gamma^{(n)}$, one 
can employ the natural orbital EOM (\ref{dynnatorb1}) in order to evaluate the 
generalized time-derivative (see eq.(\ref{gam3})) of the following anomalous local symmetry correlator 
$\phi_n({\bf x},t)\phi_n({\bf y},t)$ of the $n$-th natural orbital
\begin{align}
 i\hat\partial_t\phi_n({\bf x},t)\phi_n({\bf y},t)&\equiv
 \big[i\partial_t\phi_n({\bf x},t)\big]\phi_n({\bf y},t)-
  \phi_n({\bf x},t)\big[i\partial_t\phi_n({\bf y},t)\big]\\\nonumber
  &=\frac{1}{2}\nabla_x{\bf j}^a_n({\bf x},{\bf y},t)
  +\big(U({\bf x})-U({\bf y}) \big)\phi_n({\bf x},t)\phi_n({\bf y},t)\\\nonumber
  +N(N-1)&\sum_{p=1,p\neq n}^\infty
  \frac{I_{pn}}{\lambda_n-\lambda_p}\big(\phi_p({\bf x},t)\phi_n({\bf y},t) -
  \phi_n({\bf x},t)\phi_p({\bf y},t)\big),
\end{align}
where $I_{pn}$ denotes again the matrix elements of the collision integral in 
the natural orbital basis (see eq.(\ref{IPN})) and ${\bf j}^a_n({\bf x},{\bf y},t)$ 
refers to the following anomalous correlator current density
\begin{equation}
 {\bf j}^a_n({\bf x},{\bf y},t) = \sigma \phi_n({\bf 
x},t)\big(\nabla\phi_n\big)({\bf y},t)-\phi_n({\bf 
y},t)\big(\nabla\phi_n\big)({\bf x},t).
\end{equation}
As in the case of the canonical local symmetry correlator for an individual 
natural orbital (see eq.(\ref{onenatsymeom})), the anomalous local symmetry correlator 
$\phi_n({\bf x},t)\phi_n({\bf y},t)$ couples to its off-diagonal counterparts
$\phi_p({\bf x},t)\phi_n({\bf y},t)$, $p\neq n$, for which one can easily 
derive equations of motion, too. It is important to notice that while 
the anomalous correlators discussed in this appendix are obviously distinct 
from $\chi_i({\bf x},t) \chi_i({\bf y},t)$, their EOM reduce to 
(\ref{spc}) in the absence of interactions for the stationary situation.

However, it appears not fully satisfactory to have anomalous symmetry 
correlators and their EOM only for the individual natural 
orbitals (in analogy to eq.(\ref{onenatsymeom})). The 
anomalous counterpart of eqs.(\ref{toteom},\ref{toteomsym}), 
i.e.\ the EOM for the convex sum of the natural orbital 
anomalous symmetry correlators weighted with the respective populations, can be 
derived as follows: First, one can combine eqs.(\ref{dynnatpop},\ref{dynnatorb1})  
to an EOM for the renormalized, i.e.\
population normalized natural orbitals $\tilde\phi_n({\bf x},t)\equiv
\sqrt{\lambda_n(t)}\phi_n({\bf x},t)$ \cite{Brics}. Second, this equation is employed to evaluate
$ i\hat\partial_t\tilde\phi_n({\bf x},t)\tilde\phi_n({\bf y},t)$ and, finally, 
one can sum the result over $n$ such that one arrives at an EOM
for $\sum_{n=1}^\infty \lambda_n(t)\phi_n({\bf x},t)\phi_n({\bf y},t)$. In this 
way, one finds an alternative to eq.(\ref{gamma1diag}), which avoids the difficulties regarding diagonalizability and 
whose EOM has again the non-interacting stationary limit (\ref{spc}).

\end{document}